\documentclass[11pt]{elsart}
\usepackage{amsfonts}
\usepackage{amssymb}
\usepackage{color}
\usepackage{amssymb,amsmath}% AMS packages
\usepackage{mathrsfs} % mathcal fonts

\begin{document}
\begin{frontmatter}
\title{Constructing a type-adjustable mechanism to obtain Pareto-optimal outcomes}
\author{Haoyang Wu\corauthref{cor}}
\corauth[cor]{Corresponding author.} \ead{yangki76@163.com}
\address{Wan-Dou-Miao Research Lab, Shanghai, China.}

\begin{abstract}
In mechanism design theory, agents' types are described as their private information, and the designer may reveal some public information to affect agents' types in order to obtain more payoffs. Traditionally, each agent's private type and the public information are represented as a random variable respectively. In this paper, we propose a type-adjustable mechanism where each agent's private type is represented as a function of two parameters, \emph{i.e.}, his intrinsic factor and an external factor. Each agent's intrinsic factor is modeled as a private random variable, and the external factor is modeled as a solution of the designer's optimization problem. If the designer chooses an optimal value of external factor as public information, the type-adjustable mechanism may yield Pareto-optimal outcomes, which let the designer and each agent obtain more expected payoffs than what they would obtain at most in the traditional optimal mechanisms. As a comparison, in an auction with interdependent values, only the seller will benefit from public information which is represented as a random variable. We propose a revised version of revelation principle for type-adjustable Bayesian equilibrium. In the end, we compare the type-adjustable mechanism with other relevant models.
\end{abstract}
\end{frontmatter}

\section{Introduction}
In mechanism design theory \cite{MWG1995, Narahari2009, Serrano2004}, there are one designer and some agents\footnote{In following discussions, the designer is denoted as ``She'', and the agent is denoted as ``He''.}, and each agent is assumed to privately observe a parameter (described as ``type'') that determines his preference relation over a set of outcomes. The designer would like to implement a desired social choice function which specifies her favorite outcome for each possible profile of agents' private types. To do so, the designer constructs a mechanism which specifies each agent's strategy set ($i.e.$, the allowed actions of each agent) and an outcome function ($i.e.$, a rule for how agents' actions get turned into a social choice) \cite{MWG1995}.

Although describing an agent's type as a parameter looks reasonable in some simple cases, it should be noted that in the real world an agent's preference relation may vary with the context (\emph{e.g.}, experience, characters, emotions and so on) \cite{LS2013}. Therefore, it may not be enough to describe an agent's private type simply as a parameter. Another possible method is to represent it as a function which changes both in response to external stimuli and as a result of mental processes \cite{Hansson1995}.

Generally speaking, there are two kinds of external stimuli that an agent may be faced with: 1) \emph{Natural factors}, the value of which are chosen by the nature or taken for granted, and cannot be controlled by the designer. 2) \emph{Control factors}, the values of which are chosen by the designer\footnote{Control factors are different from the public information occurred in auctions with interdependent values \cite{Krishna2010}, where the public information sent by the seller is represented as a random variable, the value of which is chosen by nature. As a comparison, the values of control factors are chosen by the designer.}. For simplification, in the following discussions we will focus on the impact of control factors on agents' types and neglect natural factors.

During the period of a mechanism, the designer's role is different. Before announcing a mechanism, the designer looks like a powerful person since she can decide each agent's feasible strategy set and an outcome function specifying how the final outcome is yielded. However, after announcing a mechanism, the designer looks like an obedient person, since she must accept any outcome yielded by the mechanism. Actually, the designer should obey the mechanism constructed by herself, otherwise no agent is willing to believe her and participate the mechanism.

\subsection{Several ways to improve the designer's payoffs}
So far, there have been several possible ways to generalize the model of mechanism in order to improve the designer's payoffs, which are listed as follows.\footnote{In this paper, we only consider one-stage mechanism, and omit models of multi-stage mechanism or other complex circumstances (such as renegotiation, dynamic mechanism, imperfect commitment and so on).}

1) The first way is straightforward. For example, let us consider an auction, in the beginning the seller spends some cost to refresh the sold object, and hence each bidder might increase his private valuation and bid to the object. As a result, the seller may obtain more payoffs than what she might obtain without refreshing the sold object. Note that in this way the sold object has been changed before each bidder submits his bid, therefore it is trivial that the seller's benefit may be increased.

2) The designer may hold a charity auction. Engers and McManus \cite{Engers2007} proposed that agents' bids in a first-price charity auction are greater than those in a standard (non-charity) auction \cite{Myerson1981} because of the charitable benefit that winners receive from their own payments.

3) Since agents' types may change with changing circumstances (Ref \cite{Maschler2013}, Section 2.9.5), the designer may reveal public information to adjust agents' informational environments, and then induce agents to change their private types. For example, in auctions with interdependent values, the seller may have some public information that would increase the valuations of the bidders if known by them \cite{Krishna2010}. Kamenica and Gentzkow \cite{KG2011} proposed a model of Bayesian persuasion in which the sender can strategically control the receiver's information to influence her beliefs, and thus affect the actions that she takes.

\subsection{Shortcoming of representing public information as a random variable}
In this paper, we follow the above third way but do not represent public information as a random variable, which is commonly adopted in current literatures. The shortcoming of representing public information as a random variable is shown by the following example.

\textbf{Example:} Consider a revised symmetric first-price sealed-bid (\emph{FPSB}) auction, where a seller wants to sell an object to $n$ bidders. In the beginning each bidder $i$ has a private valuation $\theta_{i}$ to the sold object. The probability distribution of $\theta_{i}$ is common knowledge and identical to each other. Different from the traditional $FPSB$ model, here we assume that:\\
1) The seller chooses some non-negative value $c$ of control factor as public information to change bidders' informational environments\footnote{The control factor can be understood as an action performed by the seller to change bidders' informational environments, such as an advertisement, or a training course to improve bidders' knowledge to understand the sold object. The value of control factor corresponds to the non-negative cost of the action.}. The seller knows how each bidder's valuation to the object changes with $c$. \\
2) After learning about the value $c$, each bidder adjusts his private valuation $\theta_{i}$ and bid $b_{i}$. Hence, each agent $i$'s bid is denoted as a function of his valuation which depends on $c$, \emph{i.e.}, $b_{i}(\theta_{i}(c))$. \\
3) When the value $c$ of control factor is small, its tiny increment will yield a larger increment of each agent's bid, \emph{i.e.}, $ db_{i}(\theta_{i}(c))/dc>1$ for a small $c$. With the increasement of $c$, the ratio $db_{i}(\theta_{i}(c))/dc$ decreases.

According to footnote 4, the seller's payoffs is the gap between the winner bidder's bid and the seller's cost, \emph{i.e.},
\begin{equation*}
  \max\limits_{1\leq i\leq I} b_{i}(\theta_{i}(c)) - c.
\end{equation*}
It can be easily seen that the seller's payoffs will reach the maximum when the value of control factor reaches a critical value $c^{*}$ such that $db_{i}(\theta_{i}(c))/dc = 1$ for each bidder $i$.
Obviously, it will always be worthy for the seller to increase $c$ as long as $db_{i}(\theta_{i}(c))/dc>1$ holds.

Note that in this example, the public information (\emph{i.e.}, the value $c$ of control factor) is not random but a \emph{deterministic} solution of an optimization problem,
\begin{equation*}
  c^{*} = argmax_{c\geq0}[\max\limits_{1\leq i\leq I}b_{i}(\theta_{i}(c)) - c].
\end{equation*}
Obviously, this case should not be modeled by the traditional mechanism where the public information is represented as a random variable, the value of which is randomly chosen by the nature. This example leads to the motivation of this paper.

\subsection{Motivation}
In order to model the above-mentioned case where each agent's private type depends not only on his own character, but also on external factors chosen by the designer, we need to generalize the traditional model of mechanism to include two features: 1) Each agent's private type depends on his intrinsic factor and external factors, 2) The values of external factors can be optimally controlled by the designer, and are common knowledge.

The aim of the generalized mechanism is that: by choosing the values of control factors, the designer may have an additional power to optimally adjust the probability distribution of each agent's private type\footnote{Formal descriptions about the additional power are given in Note 1.} and then obtain Pareto-optimal outcomes, which enlarge the designer's expected payoffs and each agent's expected payoffs at the same time.

\subsection{Organizations}
This paper is organized as follows. In Section 2, we define a series of notions such as type function, type-adjustable mechanism, truthfully implementable in type-adjustable Bayesian Nash equilibrium and so on. The main result is Proposition 1, \emph{i.e.}, the type-adjustable revelation principle for Bayesian Nash equilibrium.

In Section 3, we construct a type-adjustable auction to show that the designer may obtain more expected payoffs than what she could obtain at most in the traditional optimal auction model. To the satisfaction of all, each agent's expected payoffs may be increased at the same time. In Section 4, we compare the type-adjustable mechanism with other relevant models in the literature. Section 5 draws conclusions.

\section{Definitions and main result}
Following Mas-Colell, Whinston and Green's book (\cite{MWG1995}, Section 23.B), we consider a setting with one designer and $I$ agents, each agent knows his private type but not necessarily the types of the others. The set of each agent $i$'s possible type $\theta_{i}$ is denoted as $\Theta_{i}$ ($i=1,\cdots, I$). Let $\theta=\theta_{1}\times\cdots\times\theta_{I}$, $\Theta=\Theta_{1}\times\cdots\times\Theta_{I}$.
For a set $X$ of possible outcomes and an outcome $x\in X$, the utility function of each agent $i$ with type $\theta_{i}$ is denoted as $u_{i}(x, \theta_{i}):X\times\Theta_{i}\rightarrow\mathbb{R}$, the designer's utility function is denoted as $u_{d}(x):X\rightarrow\mathbb{R}$. A social choice function (\emph{SCF}) $f:\Theta\rightarrow X$ specifies an outcome for each profile of agents' types $\theta\in\Theta$.

\textbf{\emph{Assumption 1:}} Different from the traditional model of mechanism where each agent $i$'s type $\theta_{i}$ is represented simply as a parameter, here we assume that each agent $i$'s type depends on two kinds of parameters:\\
1) Agent $i$'s intrinsic factor $\tilde{\theta}_{i}\in\Theta_{i}$: is agent $i$'s private information and is represented as a random variable\footnote{Since each agent $i$ knows his intrinsic factor $\tilde{\theta}_{i}$, $\tilde{\theta}_{i}$ is not random from his own perspective. However, because the designer does not know the exact value of $\tilde{\theta}_{i}$, what she can do is to guess it according to some subjective probability distribution. Note that the mechanism is constructed by the designer instead of agents, hence it is convenient from the designer's perspective to describe each agent's intrinsic factor as a random variable.}. The probability distribution $\tilde{\phi}_{i}(\tilde{\theta}_{i})$ is assumed to be common knowledge among the designer and other agents.\\
2) An external control factor $c\in\mathbb{R}_{+}$: non-negative and chosen by the designer. The control factor is assumed to be observable to all agents.

\textbf{Definition 1 (Type function):} \\
Each agent $i$'s type is defined by a \emph{type function} of two parameters, \emph{i.e.}, $\mu_{i}(\tilde{\theta}_{i}, c):\Theta_{i}\times\mathbb{R}_{+}\rightarrow\Theta_{i}$, which describes how the type of agent $i$ with an intrinsic factor $\tilde{\theta}_{i}$ changes with the control factor $c$. The function $\mu_{i}$ is assumed to be common knowledge.

\textbf{Definition 2 (Initial type):}\\
Consider a trivial case where the designer does not spend any effort to affect agents' types. Let this case correspond to zero value of control factor, \emph{i.e.}, $c=0$. Each agent $i$'s type in this case is defined as his \emph{initial type} $\theta^{0}_{i}$, and is equal to his intrinsic factor $\tilde{\theta}_{i}$,
\begin{equation*}
\theta^{0}_{i}\equiv\mu_{i}(\tilde{\theta}_{i}, 0)\equiv \tilde{\theta}_{i}.
\end{equation*}
The probability distribution of $\theta^{0}_{i}$ is denoted as agent $i$'s \emph{initial type distribution} $\phi^{0}_{i}(\cdot)$, and is equal to $\tilde{\phi}_{i}(\cdot)$.
Let $\phi^{0}(\theta^{0})\equiv(\phi_{1}^{0}(\theta_{1}^{0}), \cdots, \phi_{I}^{0}(\theta_{I}^{0}))$ denote the profile of all agents' initial type distributions.

\textbf{Definition 3 (Adjusted type):}\\
Consider a non-trivial case where the designer chooses a positive value of the control factor, \emph{i.e.}, $c>0$, and each agent $i$ adjusts his private type by using the type function $\mu_{i}$. Each agent $i$'s type in this non-trivial case is denoted as his \emph{adjusted type},
\begin{equation}
\theta^{c}_{i} \equiv \mu_{i}(\theta^{0}_{i}, c).
\end{equation}

Let $\theta^{c}\equiv(\theta_{1}^{c}, \cdots,\theta_{I}^{c})$, and $\mu(\theta^{0},c)\equiv(\mu_{1}(\theta^{0}_{1},c),\cdots, \mu_{I}(\theta^{0}_{I},c))$, then the profile of all agents' adjusted types $\theta^{c}=\mu(\theta^{0},c)$.
The probability distribution of $\theta^{c}_{i}$ is denoted as agent $i$'s \emph{adjusted type distribution} $\phi^{c}_{i}(\cdot)$.
Let $\phi^{c}(\theta^{c})\equiv(\phi_{1}^{c}(\theta_{1}^{c}), \cdots, \phi_{I}^{c}(\theta_{I}^{c}))$ denote the profile of all agents' adjusted type distributions.

For each $i=1, \cdots, I$, let
\begin{align*}
\theta_{-i}^{0}&\equiv(\theta_{1}^{0}, \cdots, \theta_{i-1}^{0}, \theta_{i+1}^{0}, \cdots, \theta_{I}^{0}),\\
\theta_{-i}^{c}&\equiv(\theta_{1}^{c}, \cdots, \theta_{i-1}^{c}, \theta_{i+1}^{c}, \cdots, \theta_{I}^{c}),\\
\mu_{-i}(\theta_{-i}^{0}, c)&\equiv(\mu_{1}(\theta_{1}^{0}, c), \cdots, \mu_{i-1}(\theta_{i-1}^{0}, c), \mu_{i+1}(\theta_{i+1}, c), \cdots, \mu_{I}(\theta_{I}, c)),\\
\phi_{-i}^{0}(\theta_{-i}^{0})&\equiv(\phi_{1}^{0}(\theta_{1}^{0}), \cdots, \phi_{i-1}^{0}(\theta_{i-1}^{0}), \phi_{i+1}^{0}(\theta_{i+1}^{0}), \cdots, \phi_{I}^{0}(\theta_{I}^{0})),\\
\phi_{-i}^{c}(\theta_{-i}^{c})&\equiv(\phi_{1}^{c}(\theta_{1}^{c}), \cdots, \phi_{i-1}^{c}(\theta_{i-1}^{c}), \phi_{i+1}^{c}(\theta_{i+1}^{c}), \cdots, \phi_{I}^{c}(\theta_{I}^{c})).
\end{align*}

\textbf{Note 1:} According to Assumption 1 and Definition 1, the designer knows each agent $i$'s initial type distribution $\phi^{0}_{i}(\cdot)$ and type function $\mu_{i}$. Therefore, by choosing the value of control factor $c$, the designer is able to adjust the profile of agents' initial type distributions $\phi^{0}(\theta^{0})$ to another profile $\phi^{c}(\theta^{c})$.\footnote{This is the designer's additional power discussed in Footnote 5.}

%\textbf{Note 2:} As each agent $i$'s initial type $\theta^{0}_{i}$ is his private information, the adjusted type $\theta^{c}_{i}$ is also his private information. For different value of control factor $c$, agent $i$ with initial type $\theta^{0}_{i}$ will have different adjusted type $\theta^{c}_{i}$. Consequently, before the exact value of control factor is specified, each agent $i$'s type is not determined.

%\textbf{Example 2:} Suppose there are a seller and some bidders, and the seller wants to assign a sold object to the bidder with highest bid. Assume that each worker's attitude on the job is his private type, and depends on his initial attitude (\emph{i.e.}, the intrinsic factor) and the bonus (\emph{i.e.}, the control factor chosen by the manager). Assume the manager knows the probability distribution of each worker's initial attitude, and knows how it changes with bonus. After observing the bonus, each worker $i$ ($i=1, 2$) adjusts his attitude. Therefore, the manager can freely adjust the probability distribution of each worker's private attitude by choosing a level of bonus.

\textbf{Definition 4 (Type-adjustable mechanism):} \\
Given a profile of agents' type sets $\Theta_{1}, \cdots, \Theta_{I}$, a set of outcomes $X$, a social choice function $f:\Theta_{1}\times\cdots\times\Theta_{I}\rightarrow X$ and type functions $\mu_{1}, \cdots, \mu_{I}$, a \emph{type-adjustable mechanism} is defined by $\Gamma^{c}=(S_{1}, \cdots, S_{I}, \mu_{1}, \cdots, \mu_{I}, g, c)$, which works as the following steps:\\
1) The designer chooses the value of control factor $c$, specifies a feasible strategy set $S_{i}$ for each agent $i$, and an outcome function $g: S_{1}\times\cdots\times S_{I}\rightarrow X$.\\
2) Each agent $i$ changes his type from the initial value $\theta_{i}^{0}=\tilde{\theta}_{i}$ to the adjusted value $\theta^{c}_{i} = \mu_{i}(\tilde{\theta}_{i}, c)$.\\
3) Each agent $i$ chooses a strategy $s_{i}\in S_{i}$ to perform. \\
4) The mechanism yields the final outcome $g(s_{1}, \cdots, s_{I})\in X$.

\textbf{Definition 5 (Expected payoffs of the designer and each agent):}\\
Given a social choice function $f:\Theta\rightarrow X$, a non-negative value $c$ of control factor, agents' initial types $\theta^{0}$ and initial type distributions $\phi^{0}(\theta^{0})$, \emph{the designer's adjusted expected payoffs} for $c>0$ is defined by
\begin{equation}\label{u_bar_d_c}
\bar{u}_{d}(c)\equiv E_{\theta^{c}}u_{d}(f(\theta^{c}))-c=\int_{\theta^{0}\in\Theta}u_{d}(f(\mu(\theta^{0},c)))\phi^{0}(\theta^{0})d\theta^{0}-c.
\end{equation}
Each agent $i$'s \emph{adjusted expected payoffs} for $c>0$ is defined by
\begin{equation}\label{u_bar_i_c}
\bar{u}_{i}(c)\equiv E_{\theta^{c}}u_{i}(f(\theta^{c}), \theta_{i}^{c}) =\int_{\theta^{0}\in\Theta}u_{i}(f(\mu(\theta^{0},c)), \mu(\theta_{i}^{0},c))\phi^{0}(\theta^{0})d\theta^{0}.
\end{equation}
Since $\mu(\theta^{0}, 0)=\theta^{0}$, \emph{the designer's initial expected payoffs} for $c=0$ is
\begin{equation}
\bar{u}_{d}(0)=E_{\theta^{0}}u_{d}(f(\theta^{0}))=\int_{\theta^{0}\in\Theta}u_{d}(f(\theta^{0}))\phi^{0}(\theta^{0})d\theta^{0}.
\end{equation}
Each agent $i$'s \emph{initial expected payoffs} for $c=0$ is
\begin{equation}
\bar{u}_{i}(0) = E_{\theta^{0}}u_{i}(f(\theta^{0}), \theta_{i}^{0})=\int_{\theta^{0}\in\Theta}u_{i}(f(\theta^{0}), \theta_{i}^{0})\phi^{0}(\theta^{0})d\theta^{0}.
\end{equation}
 
\textbf{\emph{Assumption 2:}} The designer's expected payoffs $\bar{u}_{d}(c)$ is a concave function with respect to the value of control factor, \emph{i.e.},
\begin{flalign*}
\begin{split}
& 1)\quad \frac{d\bar{u}_{d}(c)}{dc}\Big|_{c=0^{+}}\geq 0,\\
& 2)\quad \frac{d^{2} \bar{u}_{d}(c)}{dc^{2}}<0,\quad \text{for any } c\geq 0.\\
\end{split}&
\end{flalign*}

\textbf{Definition 6 (Optimal control value):}\\
If there exists a positive value $c^{*}>0$ satisfying the following equation,\\
\begin{equation*}
\quad\frac{d\bar{u}_{d}(c)}{dc}\Big|_{c=c^{*}}= 0,
\end{equation*}
then by Assumption 2, the designer's expected payoffs $\bar{u}_{d}(c)$ will reach its maximum at $c=c^{*}$. Here, $c^{*}$ is denoted as the \emph{optimal control value}. Obviously, $\bar{u}_{d}(c^{*})>\bar{u}_{d}(0)$ for the case of $c^{*}>0$.

\textbf{Note 2:} \\
1) If $\bar{u}_{d}(c)$ satisfies the following condition,
\begin{equation*}
\frac{d\bar{u}_{d}(c)}{dc}\Big|_{c=0^{+}}= 0,
\end{equation*}
then the designer's expected payoffs $\bar{u}_{d}(c)$ will reach its maximum at $c=0$, \emph{i.e.}, the optimal control value $c^{*}=0$, and hence $\bar{u}_{d}(c^{*})=\bar{u}_{d}(0)$.\\
2) If $\bar{u}_{d}(c)$ satisfies the following condition,
\begin{equation*}
\frac{d\bar{u}_{d}(c)}{dc}\Big|_{c=0^{+}}> 0,
\end{equation*}
then there exists a positive optimal control value $c^{*}$, and $\bar{u}_{d}(c^{*})>\bar{u}_{d}(0)$.

\textbf{Definition 7 (Type-adjustably Bayesian implementable):}\\
Given a social choice function $f$ and a profile of agents' initial type distributions $\phi^{0}(\theta^{0})$, $f$ is called \emph{type-adjustably Bayesian implementable} if the following conditions are satisfied:\\
1) There exists a positive optimal control value, \emph{i.e.}, $c^{*}>0$.\\
2) There exists a type-adjustable mechanism $\Gamma^{c^{*}}=(S_{1},\cdots,S_{I},\mu_{1},\cdots,\mu_{I},g, c^{*})$ that implements $f$ in Bayesian  equilibrium, \emph{i.e.}, there exists a strategy profile $s^{*}(\cdot)=(s^{*}_{1}(\cdot),\cdots,s^{*}_{I}(\cdot))$ such that:\\
(i) For all agent $i$, all $\theta_{i}^{c^{*}}\in\Theta_{i}$, and all $\hat{s}_{i}\in S_{i}$,
%\footnote{In formula (\ref{type-adjustable_BNE}), the expected value is computed using agents' adjusted type distributions $\phi_{-i}^{c^{*}}(\theta_{-i}^{c^{*}})$. As a comparison, in the traditional notion of Bayesian  equilibrium (Ref \cite{MWG1995}, Page 883, Definition 23.D.1), the expected value is computed using agents' initial type distributions $\phi_{-i}^{0}(\theta_{-i}^{0})$.}
\begin{equation}\label{type-adjustable_BNE}
  E_{\theta_{-i}^{c^{*}}}[u_{i}(g(s^{*}_{i}(\theta_{i}^{c^{*}}),s^{*}_{-i}(\theta_{-i}^{c^{*}})), \theta_{i}^{c^{*}})|\theta_{i}^{c^{*}}]
  \geq
  E_{\theta_{-i}^{c^{*}}}[u_{i}(g(\hat{s}_{i},s^{*}_{-i}(\theta_{-i}^{c^{*}})), \theta_{i}^{c^{*}})|\theta_{i}^{c^{*}}],
\end{equation}
\emph{i.e.}, for all agent $i$, all $\theta_{i}^{0}\in\Theta_{i}$, and all $\hat{s}_{i}\in S_{i}$,
\begin{align*}
\int_{\theta_{-i}^{0}\in\Theta_{-i}}&u_{i}(g(s^{*}_{i}(\mu_{i}(\theta_{i}^{0}, c^{*})), s^{*}_{-i}(\mu_{-i}(\theta_{-i}^{0}, c^{*}))), \mu_{i}(\theta_{i}^{0}, c^{*}))\phi_{-i}^{0}(\theta_{-i}^{0})d\theta_{-i}^{0}\\
&\geq\int_{\theta_{-i}^{0}\in\Theta_{-i}}u_{i}(g(\hat{s}_{i}, s^{*}_{-i}(\mu_{-i}(\theta_{-i}^{0}, c^{*}))), \mu_{i}(\theta_{i}^{0}, c^{*}))\phi_{-i}^{0}(\theta_{-i}^{0})d\theta_{-i}^{0}.
\end{align*}
(ii) $g(s^{*}(\theta))=f(\theta)$ for any profile of agents' types $\theta\in\Theta$.

\textbf{Note 3}: If a social choice function $f$ is type-adjustably Bayesian implementable, then according to Definition 7 and Note 2, by constructing a type-adjustable mechanism the designer will obtain more expected payoffs than what she would obtain at most in a traditional mechanism. Furthermore, if each agent $i$'s adjusted expected payoffs $\bar{u}_{i}(c)$ is an increasing function of $c$, then the type-adjustable mechanism will let each agent obtain more expected payoffs than he would obtain in the traditional mechanism. Therefore, \emph{the type-adjustable mechanism may yield Pareto-optimal outcomes better than ever before}.

% As a comparison, in an auction with interdependent values, only the seller benefits from public information: the expected revenue of the seller in a first-price auction is higher when public information is made available than when it is not (Ref \cite{Krishna2010}, Section 7.2).

%\textbf{Proof}: Since $f$ is type-adjustably Bayesian implementable, then according to Definition 7, there exists an optimal control value $c^{*}>0$ such that $\bar{u}_{d}(c^{*})$ is the designer's maximum expected payoffs with a positive value of $c$, and $\bar{u}_{d}(c^{*})>\bar{u}_{d}(0)$. $\Box$

\textbf{Definition 8 (Type-adjustable direct mechanism):}\\
Given a profile of agents' types $\Theta_{1}, \cdots, \Theta_{I}$, a set of outcomes $X$, a social choice function $f:\Theta\rightarrow X$ and type functions $\mu_{1}, \cdots, \mu_{I}$, a \emph{type-adjustable direct mechanism} is defined by $\tilde{\Gamma}^{c}=(\Theta_{1},\cdots, \Theta_{I}, \mu_{1}, \cdots, \mu_{I},f, c)$
which works as follows:\\
1) The designer chooses a non-negative value $c$ of control factor.\\
2) Each agent $i$ reports a type $\theta_{i}\in\Theta_{i}$, and $\theta_{i}$ does not need to be equal to his initial type $\tilde{\theta}_{i}$.\\
3) The mechanism yields an outcome $f(\theta^{c}_{1}, \cdots, \theta^{c}_{I})$, where $\theta^{c}_{i}=\mu_{i}(\theta_{i}, c)$.

\textbf{Note 4}: In a type-adjustable direct mechanism, each agent $i$'s expected utility $\bar{u}_{i}(c)=E_{\theta_{i}^{c}}u_{i}(f(\theta_{1}^{c}, \cdots, \theta_{I}^{c}), \theta_{i}^{c})$ is based on agents' adjusted type distributions $\phi^{c}(\theta^{c})$, not on agents' initial type distributions $\phi^{0}(\theta^{0})$. Therefore, the revelation principle and the notion of incentive compatibility should be revised for the case of type-adjustable mechanisms.

\textbf{Definition 9 (Truthfully implementable in type-adjustable Bayesian Nash equilibrium):}\\
A social choice function $f(\cdot)$ is \emph{truthfully implementable in type-adjustable Bayesian Nash equilibrium} (or \emph{type-adjustably Bayesian incentive compatible}) if $s^{*}_{i}(\theta_{i})=\theta_{i}$ (for all
$\theta_{i}\in\Theta_{i}$ and $i=1, \cdots, I$) is a Bayesian Nash equilibrium of a type-adjustable direct mechanism $\tilde{\Gamma}^{c^{*}}=(\Theta_{1},\cdots, \Theta_{I}, \mu_{1}, \cdots, \mu_{I}, f, c^{*})$ That is, if for all $i=1, \cdots, I$, $\theta_{i}^{c^{*}}\in\Theta_{i}$, $\hat{\theta}_{i}\in\Theta_{i}$,
\begin{equation}\label{type-adjustable_BIC}
  E_{\theta_{-i}^{c^{*}}}[u_{i}(f(\theta_{i}^{c^{*}},\theta_{-i}^{c^{*}}), \theta_{i}^{c^{*}})|\theta_{i}^{c^{*}}]
  \geq
  E_{\theta_{-i}^{c^{*}}}[u_{i}(f(\hat{\theta}_{i},\theta_{-i}^{c^{*}}), \theta_{i}^{c^{*}})|\theta_{i}^{c^{*}}],
\end{equation}
\emph{i.e.}, for all $i=1, \cdots, I$, $\theta_{i}^{0}\in\Theta_{i}$, $\hat{\theta}_{i}\in\Theta_{i}$,
\begin{align*}
\int_{\theta_{-i}^{0}\in\Theta_{-i}}&u_{i}(f(\mu_{i}(\theta_{i}^{0}, c^{*}), \mu_{-i}(\theta_{-i}^{0}, c^{*})), \mu_{i}(\theta_{i}^{0}, c^{*}))\phi_{-i}^{0}(\theta_{-i}^{0})d\theta_{-i}^{0}\\
&\geq\int_{\theta_{-i}^{0}\in\Theta_{-i}}u_{i}(f(\hat{\theta}_{i}, \mu_{-i}(\theta_{-i}^{0}, c^{*})), \mu_{i}(\theta_{i}^{0}, c^{*}))\phi_{-i}^{0}(\theta_{-i}^{0})d\theta_{-i}^{0}.
\end{align*}

\textbf{Proposition 1 (Type-adjustable revelation principle for Bayesian Nash equilibrium):}\\
If a social choice function is type-adjustably Bayesian implementable, then it is also truthfully implementable in type-adjustable Bayesian Nash equilibrium.\\
\textbf{Proof:} The proof is straightforward.

\section{A type-adjustable first-price sealed-bid auction}
Following the first-price sealed-bid (\emph{FPSB}) auction given in MWG's book (Ref \cite{MWG1995}, Example 23.B.5, Page 865), in this section we will construct a type-adjustable \emph{FPSB} auction which may yield Pareto-optimal outcomes, \emph{i.e.}, both the seller and each bidder may obtain more expected payoffs than what they might obtain at most from the traditional \emph{FPSB} auction.

\subsection{Model settings}
Suppose that there are one seller and two bidders. Each bidder's private type is his valuation to a sold object, which depends on his initial valuation and an external control factor chosen by the seller. Each bidder $i$'s initial valuation $\theta_{i}^{0}$ is drawn independently from the uniform distribution on $[0,1]$. Let $\beta>0$ be a coefficient describing the significance of the impact of control factor on each bidder's valuation to the sold object. The greater the value of $\beta$ is, the more significant the impact of control factor on each bidder's valuation to the sold object will be. $\beta$ is assumed to be common knowledge.

The type-adjustable \emph{FPSB} auction works as the following steps.\\
\emph{Step 1}: The seller chooses a non-negative value $c$ of control factor, which is observable to two bidders.\\
\emph{Step 2}: After observing the value $c$ of control factor, each bidder $i$ adjusts his initial type according to a concave type function, which is common knowledge,
\begin{equation}\label{theta_i}
\theta_{i}^{c}= (1+\beta\sqrt{c})\theta_{i}^{0}.
\end{equation}\\
\emph{Step 3}: Each bidder $i$ with adjusted type $\theta_{i}^{c}$ submits a sealed bid $b_{i}\geq 0$ to the seller.\\
\emph{Step 4}: The bidder who submits the higher bid wins the object, and pays money equal to his bid to the seller.

Let $\theta=(\theta_{1},\theta_{2})$ denote a profile of two bidders' types. Consider the following social choice function
\begin{equation}\label{SCF f}
f(\theta)=(y_{1}(\theta), y_{2}(\theta), y_{d}(\theta), t_{1}(\theta), t_{2}(\theta), t_{d}(\theta)),
\end{equation}
in which
\begin{align*}
&y_{1}(\theta)=1, \quad \text{if }\theta_{1}\geq\theta_{2};\quad =0 \text{ if }\theta_{1}<\theta_{2}\\
&y_{2}(\theta)=1, \quad \text{if }\theta_{1}<\theta_{2};\quad =0 \text{ if }\theta_{1}\geq\theta_{2}\\
&y_{d}(\theta)=0, \quad\text{for all }\theta\in\Theta\\
&t_{1}(\theta)= - \theta_{1} y_{1}(\theta)/2\\
&t_{2}(\theta)= - \theta_{2}y_{2}(\theta)/2\\
&t_{d}(\theta)= [\theta_{1}y_{1}(\theta)+\theta_{2} y_{2}(\theta)]/2.
\end{align*}
The subscript $``d"$ stands for the seller (\emph{i.e.}, the designer of this auction), and the subscript $``1",``2"$ stands for the bidder 1 and bidder 2 respectively. $y_{i}=1$ means that bidder $i$ gets the object, $t_{i}$ denotes bidder $i$'s payment to the seller, $t_{d}$ denotes the sum of two bidders' payment to the seller.

\subsection{The SCF f is Bayesian  implementable}
Let us investigate whether the social choice function $f(\theta)$ is Bayesian  implementable. We will look for a Bayesian  equilibrium in which each bidder $i$'s strategy $b_{i}(\cdot)$ takes the form $b_{i}(\theta_{i}^{c})=\alpha_{i}\theta_{i}^{c}=\alpha_{i}(1+\beta\sqrt{c})\theta_{i}^{0}$ for $\alpha_{i}\in[0,1]$.

Suppose that bidder 2's strategy has this form, then for each possible $\theta_{1}^{c}$, bidder 1's problem is to solve the following problem to find an optimal strategy $b_{1}$:
\begin{equation}\label{bidder1}
  \max\limits_{b_{1}\geq 0}(\theta_{1}^{c} - b_{1}) \text{Prob}(b_{2}(\theta_{2}^{c})\leq b_{1}).
\end{equation}
Because bidder 2's highest possible bid is $\alpha_{2}(1+\beta\sqrt{c})$  when $\theta_{2}^{0}=1$, it is evident that bidder 1's bid $b_{1}$ should not be greater than $\alpha_{2}(1+\beta\sqrt{c})$, \emph{i.e.}, $b_{1}\leq\alpha_{2}(1+\beta\sqrt{c})$.

Note that $\theta_{2}^{0}$ is uniformly distributed on $[0,1]$, and $b_{2}(\theta_{2}^{c})=\alpha_{2}(1+\beta\sqrt{c})\theta_{2}^{0}\leq b_{1}$ means that $\theta_{2}^{0}\leq b_{1}/[\alpha_{2}(1+\beta\sqrt{c})]$. Thus,
\begin{equation*}
\text{Prob}(b_{2}(\theta_{2}^{c})\leq b_{1}) = \text{Prob}(\theta_{2}^{0}\leq b_{1}/[\alpha_{2}(1+\beta\sqrt{c})]) = \frac{b_{1}}{\alpha_{2}(1+\beta\sqrt{c})}.
\end{equation*}
Now we can rewrite bidder 1's problem (formula \ref{bidder1}) as:
\begin{equation*}
  \max\limits_{0\leq b_{1}\leq \alpha_{2}(1+\beta\sqrt{c})}\frac{(\theta_{1}^{c} - b_{1}) b_{1}}{\alpha_{2}(1+\beta\sqrt{c})}
\end{equation*}
The solution to this maximum problem is
\begin{equation*}
  b_{1}^{*}(\theta_{1}^{c})
  =\begin{cases}
     \theta_{1}^{c}/2,& \text{if } \theta_{1}^{0}/2\leq\alpha_{2}\\
      \alpha_{2}(1+\beta\sqrt{c}),& \text{if } \theta_{1}^{0}/2>\alpha_{2}
   \end{cases}.
\end{equation*}
Similarly,
\begin{equation*}
  b_{2}^{*}(\theta_{2}^{c})
  =\begin{cases}
     \theta_{2}^{c}/2,& \text{if } \theta_{2}^{0}/2\leq\alpha_{1}\\
      \alpha_{1}(1+\beta\sqrt{c}),& \text{if } \theta_{2}^{0}/2>\alpha_{1}
   \end{cases}.
\end{equation*}
Let $\alpha_{1}=\alpha_{2}=1/2$, we see that the strategies $ b_{i}^{*}(\theta_{i}^{c}) = \theta_{i}^{c}/2=(1+\beta\sqrt{c})\theta_{i}^{0}/2$ for $i=1, 2$ constitute a Bayesian equilibrium for this type-adjustable \emph{FPSB} auction.

Thus, the \emph{SCF} $f$ is implemented in Bayesian equilibrium by the type-adjustable \emph{FPSB} auction. Hence, $f$ is Bayesian implementable.

\subsection{The SCF f is type-adjustably Bayesian implementable}
By Definition 5, the seller's expected payoffs for a positive $c$ is :
\begin{align*}
 \bar{u}_{d}(c)&= E_{\theta^{c}}u_{d}[f(\theta^{c})]-c\\
                    &= E_{\theta^{c}}[t_{d}(\theta^{c})]-c\\
                    &= E_{\theta^{c}}[\theta_{1}^{c}y_{1}(\theta^{c})+\theta_{2}^{c}y_{2}(\theta^{c})]/2 - c\\
                    &= (1+\beta\sqrt{c})E[\theta_{1}^{0}y_{1}(\theta^{0})+\theta_{2}^{0}y_{2}(\theta^{0})]/2 - c.
\end{align*}
By Appendix, the seller's expected payoffs for $c=0$ is
\begin{equation*}
\bar{u}_{d}(0)=E[\theta_{1}^{0}y_{1}(\theta^{0})+\theta_{2}^{0}y_{2}(\theta^{0})]/2=1/3.
\end{equation*}
Hence, $\bar{u}_{d}(c)=(1+\beta\sqrt{c})/3 - c$, and it satisfies Assumption 2. The seller's problem is to find an optimal control value to maximize $\bar{u}_{d}(c)$,
\begin{equation*}
  \max\limits_{c\geq0}[(1+\beta\sqrt{c})/3 - c].
\end{equation*}
Obviously, the optimal control value is $c^{*}=\beta^{2}/36>0$. According to Section 3.2, the strategies $ b_{i}^{*}(\theta_{i}^{c^{*}}) = \theta_{i}^{c^{*}}/2= (1+\beta\sqrt{c^{*}})\theta_{i}^{0}/2$ (for $i=1, 2$) constitute a Bayesian equilibrium for the type-adjustable \emph{FPSB} auction. By Definition 7, the social choice function $f$ is type-adjustably Bayesian implementable.

\subsection{The seller's expected payoffs for a positive $c^{*}$}
Since $c^{*}=\beta^{2}/36$, the seller's expected payoffs for a positive $c^{*}$ is:
\begin{equation*}
  \bar{u}_{d}(c^{*})= (1+\beta\sqrt{c^{*}})/3 - c^{*} = \frac{1}{3}(1+\frac{\beta^{2}}{12}).
\end{equation*}
Hence, if $\beta>\sqrt{3}$, then $\bar{u}_{d}(c^{*})>5/12$. Note that the seller's maximum expected payoffs in the traditional optimal auction with two bidders is $5/12$ (Ref \cite{Krishna2010}, Page 23, the ninth line from the bottom). Therefore, if $\beta>\sqrt{3}$, then by choosing the optimal control value $c^{*}=\beta^{2}/36$, \emph{the seller can obtain more expected payoffs than what she could obtain at most in the traditional optimal auction model}.

\subsection{Each bidder's ex ante expected payoffs for a positive $c^{*}$}
Now we consider each bidder's \emph{ex ante} expected payoffs when the seller chooses the optimal control value $c^{*}=\beta^{2}/36$. By appendix, for the case of two bidders, the expected payoffs of the winner bidder is denoted as follows:
\begin{align*}
E[\theta_{winner}^{c^{*}} - b_{winner}^{*}(\theta_{winner}^{c^{*}})] &= E[\theta_{winner}^{c^{*}}/2] =(1+\beta\sqrt{c^{*}})E[\theta_{winner}^{0}]/2\\
&=(1+\beta\sqrt{c^{*}})E[\theta_{1}^{0}y_{1}(\theta^{0})+\theta_{2}^{0}y_{2}(\theta^{0})]/2\\
&=\frac{1}{3}+\frac{\beta^{2}}{18}.
\end{align*}
Note that the expected payoffs of the loser bidder is zero. Since the two bidders are symmetric, then each of them has the same probability $1/2$ to be the winner bidder. Therefore, each bidder's \emph{ex ante} expected payoffs is half of the winner's expected payoffs, \emph{i.e.}, $1/6+\beta^{2}/36$.

\subsection{Comparison with optimal auction model}
Let us recall the traditional optimal auction model, \emph{i.e.}, the standard first-price auction model with reserve price (Ref \cite{Krishna2010}, Page 21). There is one object for sale, and $N$ potential buyers are bidding for the object. Let $r>0$ be the reserve price, and $[r, \omega]$ be the interval of each bidder $i$'s valuation which is independently and identically distributed according to an increasing distribution function $F$. Fix a bidder, $G$ denotes the distribution function of the highest valuation among the rest remaining bidders. According to Krishna's book (Ref \cite{Krishna2010}, Page 22, Line 13), the \emph{ex ante} expected payment of a bidder is
\begin{equation}\label{f41}
  r(1-F(r))G(r)+\int_{r}^{\omega}y(1-F(y))g(y)dy.
\end{equation}
For the case of two bidders with valuation range $[r, 1]$ and uniform distribution,
\begin{align*}
  &F(r)=r,\text{ }G(r)=r, \text{ }\omega=1,\\
  &F(y)=y, \text{ }g(y)=1, \text{for any } y\in[r, 1].
\end{align*}
By Ref \cite{Krishna2010} (Page 23), when each of two bidder's valuation to the object is uniformly distributed on interval $[0, 1]$, the optimal reserve price $r^{*}=1/2$. Therefore, each bidder's \emph{ex ante} expected payment in formula (\ref{f41}) is
\begin{align*}
r^{*}&(1-r^{*})r^{*}+\int_{r^{*}}^{1}y(1-y)dy\\
&=\frac{1}{8}+\int_{\frac{1}{2}}^{1}y(1-y)dy=\frac{5}{24}.
\end{align*}
Since the optimal reserve price is $1/2$, then each bidder's valuation to the object is uniformly distributed on interval $[1/2, 1]$. Hence, each bidder's expected valuation is the middle point of interval $[1/2, 1]$, $i.e.$, $3/4$.

Consequently, in the traditional optimal auction, each bidder's maximal \emph{ex ante} expected payoffs is his expected valuation $3/4$ minus his \emph{ex ante} expected payment $5/24$, $i.e.$,
\begin{equation}
 \frac{3}{4}-\frac{5}{24}=\frac{13}{24}.
\end{equation}

As specified in Section 3.5, in the type-adjustable first-price sealed-bid auction, when the seller chooses the optimal control value $c^{*}=\beta^{2}/36$, each bidder's \emph{ex ante} expected payoffs will be $1/6+\beta^{2}/36$.

Obviously, if the coefficient $\beta>\sqrt{27/2}$, then $(1/6+\beta^{2}/36)>13/24$, \emph{i.e.}, each bidder's \emph{ex ante} expected payoffs obtained in the type-adjustable \emph{FPSB} auction will be greater than the maximal \emph{ex ante} expected payoffs $13/24$ obtained in the traditional optimal auction.

According to Section 3.4, if the coefficient $\beta>\sqrt{3}$, then by choosing the optimal control value $c^{*}=\beta^{2}/36$, the seller can obtain more expected payoffs than what she could obtain at most in the traditional optimal auction model.

To sum up, if the coefficient $\beta>\sqrt{27/2}$, then for the social choice function specified by formula (\ref{SCF f}) in Section 3.1, not only the seller but also each bidder can obtain more payoffs from the type-adjustable mechanism than what they could obtain at most from the traditional optimal auctions. This is a Pareto-optimal solution.

\section{Comparisons with related models}
\subsection{Auctions with interdependent values}
In the auction with interdependent values, the public information sent by the seller is represented as a random variable, and each bidder's initial private valuation to the sold object is relevant to each other.

Contrarily, in the type-adjustable mechanism, the public information sent by the designer is represented by a deterministic solution of an optimization problem, not by a random variable. Each bidder's initial valuation is irrelevant to each other, and depends on his own intrinsic factor (\emph{i.e.}, initial valuation) and the control factor (\emph{i.e.}, public information). The type-adjustable mechanism may yield Pareto-optimal outcomes, whereas there is no such conclusion in the auction with interdependent values.

%\subsection{Dynamic mechanism design}
%Bergemann and V$\ddot{a}$lim$\ddot{a}$ki \cite{BV2019} proposed that agents' types may change in a nontrivial manner across periods of a dynamic mechanism. In Bergemann and V$\ddot{a}$lim$\ddot{a}$ki's model, each agent's private type is controllable, and the type transition function of each agent is common knowledge.
%\emph{Note:} Dynamic mechanism and the type-adjustable mechanism are different: 1) The game yielded by the type-adjustable mechanism is a one-stage game, not a multistage game as specified in the literature of dynamic mechanisms \cite{BV2019} \cite{Pavan2014}. 2) The type function in the type-adjustable mechanism is a  function, whereas the type transition function of each agent $i$ in dynamic mechanism is a stochastic function following a Markov process.

\subsection{Signaling games}
Since the designer performs a costly action as an open signal to agents, some one may consider the type-adjustable mechanism to be similar with the signaling game model. However, the two models are different:

1) In the signaling games (Ref \cite{Fudenberg1990}, Section 8.2.1, Page 324), there are one leader (Sender) and one follower (Receiver). The Sender has private information about her type, and the Receiver has no private information. Before the game begins, it is common knowledge that the Receiver has prior beliefs about the Sender's private type. The Sender moves first, and the Receiver observes the Sender's action (\emph{i.e.}, signal), then updates his belief about the Sender's type and chooses his own action. \\
The equilibrium of the signaling game is the perfect Bayesian equilibrium, which is a set of strategies and beliefs such that at any stage of the game, strategies are optimal given the beliefs, and beliefs are obtained from equilibrium strategies and observed actions using Bayes' rule.

2) In the type-adjustable mechanism, the leader (\emph{i.e.}, the designer of a mechanism, or the seller of an auction) has no private information, and the followers (\emph{i.e.}, the agents of an mechanism, or the bidders of an auction) have private information (\emph{i.e.} private types, or private valuations to the sold object). The leader moves first (\emph{i.e.}, sends a costly signal), then the followers observe the leader's action and choose their own actions (\emph{i.e.}, perform their strategies, or submits their bids). At last, the mechanism yields the outcome according to the outcome function. \\
The equilibrium of the type-adjustable mechanism is the type-adjustable Bayesian equilibrium defined by formula (\ref{type-adjustable_BNE}).

\subsection{Bayesian persuasion}
Kamenica and Gentzkow \cite{KG2011} proposed a model of Bayesian persuasion. The model consists of a player called Sender and a player called Receiver. Each of them has a utility function depending on the Receiver's action and the state of the world. A signal is a map from the state to the distribution over signal realizations. Receiver uses Bayes' rule to update his belief from the prior to the posterior, and then chooses an action that maximizes his expected utility. Given this behavior by Receiver, Sender solves an optimization problem to maximize her expected utility.

\emph{Note:} The distinctions between Bayesian persuasion and the type-adjustable mechanism are as follows: 1) In Bayesian persuasion, there is only one player ``Receiver'' besides the Sender. The signal chosen by the Sender is a map from a state to the distribution over signal realizations. 2) In the type-adjustable mechanism, there are multiple players besides the designer. There is no such notion of ``state of the world'', and the control factor chosen by the designer is a solution of an optimization problem.\\

\subsection{Information design}
According to Kamenica's descriptions \cite{Kamenica2019}, information design is similar to Bayesian persuasion. However, information design is used more when the designer is a social planner and there are multiple interacting receivers, and Bayesian persuasion is used more when the designer is one of the players in the game and there is a single receiver. Generally speaking, the information designer chooses one decision rule among decision rules, which encode the information that the receivers obtain about the realized state of the world, the types and actions of the other players.

\emph{Note:} In the model of information design, the information sent to the players is modeled as a rule which obeys a probability distribution. Contrarily, in the type-adjustable mechanism, there is no such decision rule and notion of ``state of the world''. The public information is a solution of the designer's optimization problem.

\subsection{Information structures in optimal auctions}
Bergemann and Pesendorfer \cite{BP2007} proposed a model of information structures in optimal auctions, where the seller may control bidders' information structures which generate the bidders' private information. The optimal information structures exhibit a number of properties: (i) information structures can be represented as monotone partitions, (ii) the cardinality of each partition is finite, (iii) the partitions are asymmetric across agents. Obviously, this optimal information structure is different from the type-adjustable mechanism.

\section{Conclusions}
In this paper, we propose a type-adjustable mechanism where each agent's private type is represented as a function of two kinds of parameters, \emph{i.e.}, each agent's intrinsic factor, and a control factor chosen by the designer. The control factor is modeled as a deterministic solution of the designer's optimization problem, not as a random variable. The advantage of the type-adjustable mechanism is that it may yield Pareto-optimal outcomes, beneficial not only to the designer but also to all agents. The main result is the type-adjustable revelation principle for Bayesian Nash equilibrium.

The example in Section 3 gives the following results: \\
1) For a type-adjustably Bayesian implementable social choice function, the seller may obtain more expected payoffs than her initial expected payoffs.\\
2) As shown in Section 3, the seller may breakthrough the limit of expected payoffs which she could obtain at most in the traditional optimal auction model:\\
$\bullet$ If $\beta>\sqrt{3}$, then by choosing the optimal control value $c^{*}=\beta^{2}/36$, the seller can obtain more expected payoffs than the maximum expected payoffs $5/12$ yielded by the traditional optimal auction. \\
$\bullet$ If $\beta>\sqrt{27/2}$, then each bidder's \emph{ex ante} expected payoffs $1/6+\beta^{2}/36$ will be greater than the corresponding value $13/24$ obtained in the traditional optimal auction. Put differently, every member in the type-adjustable auction may benefit from the seller's optimal signal, and hence this is Pareto-efficient.

\section*{Appendix}
As specified in Section 3, $\theta_{1}^{0}$ and $\theta_{2}^{0}$ are drawn independently from the uniform distribution on $[0,1]$. Let $Z$ be a random variable $Z=\theta_{1}^{0}y_{1}(\theta^{0})+\theta_{2}^{0}y_{2}(\theta^{0})$.
\begin{equation*}
  f_{\theta_{1}^{0}}(z)
  =\begin{cases}
      0,& z<0\\
      1,& z\in[0,1]\\
      0,& z>1
   \end{cases}.
\end{equation*}
\begin{equation*}
  F_{\theta_{1}^{0}}(z)=Prob\{\theta_{1}^{0}\leq z\}
  =\begin{cases}
      0,& z<0\\
      z,& z\in[0,1]\\
      1,& z>1
   \end{cases}.
\end{equation*}
\begin{equation*}
  F_{Z}(z) =  [F_{\theta_{1}^{0}}(z)]^2
  =\begin{cases}
      0,& z<0\\
      z^2,& z\in[0,1]\\
      1,& z>1
   \end{cases}.
\end{equation*}
Therefore,
\begin{equation*}
  f_{Z}(z)
  =\begin{cases}
      0,& z<0\\
      2z,& z\in[0,1]\\
      0,& z>1
   \end{cases}.
\end{equation*}
As a result,
\begin{equation*}
  E(Z)=\int_{0}^{1}z\cdot 2zdz=\int_{0}^{1}2z^2dz=2/3.
\end{equation*}
Therefore, $E[\theta_{1}^{0}y_{1}(\theta^{0})+\theta_{2}^{0}y_{2}(\theta^{0})]/2 = 1/3$.

By formula (\ref{SCF f}), the seller's initial expected payoffs is the sum of two bidders' payment to the seller when the cost is zero, $\bar{u}_{d}(0)=E[\theta_{1}^{0}y_{1}(\theta^{0})+\theta_{2}^{0}y_{2}(\theta^{0})]/2=1/3$.

%**********************************************************************

\end{document}